\documentclass[aps,prl,epsf,twocolumn]{revtex4}
\usepackage{graphicx,epsf}
\usepackage{dcolumn}
\usepackage{bm}

\begin{document}



\def\appConv{Appendix~A}
\def\appBeer{Appendix~B}
\def\appHValid{Appendix~C}

\def\HfO2{{HfO$_2$}}
\def\SiO2{{SiO$_2$}}
\def\Al2O3{{Al$_2$O$_3$}}
\def\Ti2O3{{Ti$_2$O$_3$}}
\def\k{{$\kappa$~}}
\def\ks{{$\kappa_s$~}}
\def\ksten{{$\tensor{\kappa}_s$~}}
\def\kinf{{$\kappa_\infty$~}}
\def\kion{{$\kappa_{ion}$~}}
\def\kionten{{$\tensor{\kappa}_{ion}$~}}
\def\kinften{{$\tensor{\kappa}_\infty$~}}
\def\cm1{{cm$^{-1}$}}
\def\zstar{{$Z^{\star}$~}}

\def\comment#1{{\large\textsl{#1}}}
\def\degree {{$^\circ$}}
\def\degrees{{$^\circ$}}
\def\eq#1{{Eq.~(\ref{eq:#1})}}
\def\fig#1{{Fig.~\ref{fig:#1}}}
\def\sec#1{{Sec.~\ref{sec:#1}}}
\def\inv{^{-1}}
\def\micron {\hbox{$\mu$m}}
\def\microns{\micron}
\def\Ref#1{{Ref.~\onlinecite{#1}}}  
\def\tab#1{{Table~\ref{tab:#1}}}
\def\tauOne{\tau^{(1)}}
\def\tVec{\hbox{\bf t}}
\def\thetaDet{\theta_{DET}}

\def\qvec{{\vec q}}
\def\pvec{{\vec p}}
\def\Avec{{\vec A}}
\def\qhat{{\hat q}}
\def\qperphat{{\hat q_\perp}}
\def\ekpq{{E_{\kvec+\qvec}}}
\def\ek{{E_{\kvec}}}
\def\Omegabar{{\bar\Omega}}
\def\omegabar{{\bar\omega}}
\def\omegap{{\omega_p}}
\def\kf{{k_F}}
\def\kappaf{{\kappa_F}}
\def\mone{{-1}}
\def\re{{\rm{Re\,}}}
\def\im{{\rm{Im\,}}}
\def\twopi{{2 \pi}}
\def\wpm{w_\pm}
\def\FWlindhard{Appendix~A}
\def\lindhardTrans{Appendix~B}
\def\ftr{{f^{tr}}}
\def\PN{Pines and Nozi{\`e}res}
\def\Bohm{B{\"o}hm}
\def\Nifosi{Nifos{\'\i}}
\def\prin{{\cal P}}

\def\imagOmegaSq{the Appendix}
\def\epsTensor{{\buildrel \leftrightarrow \over \epsilon}}
\def\chiTensor{{\buildrel \leftrightarrow \over \chi}}
\def\idenTensor{{\buildrel \leftrightarrow \over I}}
\def\epsTrans{{\epsilon^{(t)}}}
\def\epsLong{{\epsilon^{(\ell)}}}
\def\epsTransInv{{\epsilon^{(t)-1}}}
\def\epsLongInv{{\epsilon^{(\ell)-1}}}

\def\MvecA{{M^{(\vec A)}}}
\def\MdivA{{M^{(\nabla \cdot \vec A)}}}
\def\Mphi{{M^{(\phi)}}}
\def\backGrad{{\buildrel \leftarrow \over \nabla}}
\def\backMom{i \hbar \backGrad}

\def\half{{1/2}}
\def\minusHalf{{-1/2}}
\def\threeHalves{{3/2}}
\def\minusThreeHalves{{-3/2}}

\newenvironment{bulletList}{\begin{list}{$\bullet$}{}}{\end{list}}

\title{Structure of Periodic Crystals and Quasicrystals in Ultrathin Film Ba-Ti-O}

\author{Eric Cockayne\footnote{Electronic address: eric.cockayne@nist.gov}}

\affiliation{
Material Measurement Laboratory, 
National Institute of Standards and Technology, 
Gaithersburg, Maryland 20899 USA}

\author{Marek Mihalkovi\v c}

\affiliation{Institute of Physics, Slovak Academy of Sciences,
84511 Bratislava, Slovakia}



\author{Christopher L. Henley\footnote{Deceased 29 June 2015}},

\affiliation{Department of Physics, Cornell University,
Ithaca, New York 14850 USA}

\date{\today}

\begin{abstract}


 We model the remarkable thin-film Ba-Ti-O structures formed by heat treatment of an 
initial perovskite BaTiO$_3$ thin film on a Pt(111) surface.   
All structures contain a rumpled Ti-O network with all Ti threefold coordinated with O, 
and with Ba occupying the larger. mainly Ti$_7$O$_7$, pores.  The quasicrystal structue
is a simple decoration of three types of tiles:  square, triangle and
30$^{\circ}$ rhombus, with edge lengths 6.85~\AA, joined 
edge-to-edge  in a quasicrystalline pattern; observed periodic crystals in  
ultrathin film Ba-Ti-O are built from these and other tiles.
Simulated STM images reproduce the patterns seen experimentally, 
and identify the bright protrusions as Ba atoms.
The models are consistent with all experimental observations.



\end{abstract}


\maketitle
\thispagestyle{empty}


 Quasicrystals have fascinated the materials world since their
discovery\cite{Shechtman84} due to their noncrystallographic
symmetries and quasiperiodic translational order.
The first quasicrystal reported was in an Al-Mn alloy.
While numerous families of quasicrystals have since been
found\cite{Tsai13}, until recently, all known physically realized 
quasicrystals were either intermetallic alloys or 
soft matter systems.\cite{Zeng04,Lifshitz07}
This changed in 2013 with the report of a quasicrystal {\em oxide}
by  F\"orster {\it et al.}\cite{Foerster13} in a thin film Ba-Ti-O
structure created by a multistep heat treatment of perovskite
BaTiO$_3$ deposited on a Pt(111) surface, among other, periodic, 
structures.  The quasicrystalline structure is identified as a dodecagonal 
quasicrystal by its twelvefold electron diffraction pattern, and by 
scanning tunneling microscopy (STM) images that show bright protrusions separated by a distance
of about 6.85~\AA, in positions characteristic of a dodecagonal
rectangle-triangle- 30\degree~rhombus tiling\cite{Gaehler88}.
Additional experimental details distinguish the phase from perovskite
BaTiO$_3$.   The Ti ions appear to have charge 3+ via photoemission spectroscopy.
The overall stoichiometry (including both the quasicrystalline layer
and BaTiO$_3$ islands that remain after the heat treatment used
to create the quasicrystal) is Ba$_{0.9}$TiO$_{2.8}$.
While these observations give intriguing information about the 
nature of the quasicrystalline structure, a complete structure
determination is lacking.
In this paper, we give tiling decoration models for 
the atomic structure of thin film quasicrystalline Ba-Ti-O and its
related periodic crystalline structures, fully consistent with 
experimental observations.  


\begin{figure}
\includegraphics[width=236pt]{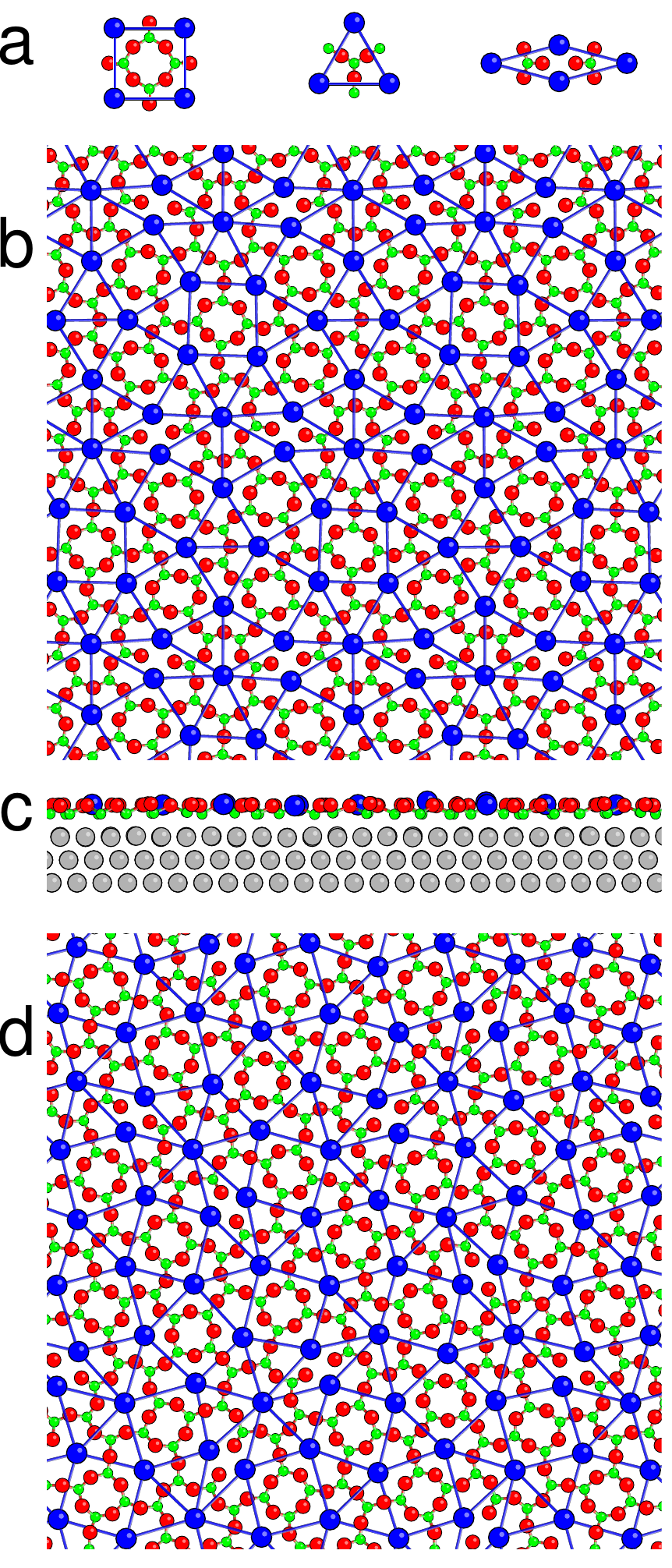}
\caption{(a) Decoration of square, triangle, and 30\degree~rhombus (rhomb) tiles.
Ba atoms are large blue; Ti small green, and oxygen medium red.
(b-d) DFT-relaxed structures of Ba-Ti-O
quasicrystalline approximants on a Pt-111 surface. Pt atoms gray.
(a) 25.6~\AA~approximant; top view.
(b) 25.6~\AA~approximant; front view.
(c) 49.4~\AA~approximant; top view.}
\label{fig:relax}
\end{figure}


 Our tile decoration model for the quasicrystal is shown in \fig{relax}(a).  The Ba, Ti, and O atoms overlay
the topmost Pt layer and project into a plane as shown.  Ba atoms occupy the
vertices of the tiles.  All Ti atoms are 3-fold coordinated with oxygen.
Most oxygen atoms are twofold coordinated with Ti except for those inside
the 30\degree~rhombi (henceforth ``rhombs").
The tiles join edge-to-edge.  
The atomic positions of the ideal tile structures match perfectly for
triangles to join rhombs and squares; the other tile combinations
require merger of the tile O positions near the shared edge.

Squares, triangles, and rhombs can be combined to produce 
numerous periodic and aperiodic tilings.
The square-triangle-rhomb tiling pattern formed by connecting 
the bright spots in the STM images\cite{Foerster13} bears a striking resemblance
to the ``$C_a$" tiling pattern found by G\"ahler\cite{Gaehler88},
using projection from a higher dimensional space with a concave
acceptance region.
This tiling (and its sufficiently large periodic approximants),
contains only square-triangle, triangle-triangle, and triangle-thin
rhomb edges.
The G\"ahler tiling has squares, triangles, and rhombs
in the relative frequency $\sqrt{3} + 1 :  2 \sqrt{3} + 4 : 1$.
Assuming that this tiling is the appropriate tiling for quasicrystalline
Ba-Ti-O, our tile decoration model gives a Ba:Ti:O ratio of
$(\sqrt{3} - 1)/2 : 1 : (3 \sqrt{3} + 1)/4$ or a stoichiometry of
approximately Ba$_{0.37}$TiO$_{1.55}$.  
From X-ray photoelectron spectroscopy measurements\cite{Foerster13},
it was deduced that there were both Ti$^{4+}$ and Ti$^{3+}$ ions 
in their sample in the ratio 5:1.  Assuming that all the Ti ion
the quasicrystalline film have charge 3+, all the Ti in the
BaTiO$_3$ islands charge 4+, and that the quasicrystalline film has the
composition found in our model, the overall composition rounds
to Ba$_{0.9}$TiO$_{2.8}$, exactly as determined experimentally.\cite{Foerster13}

 To test the stability of our model, refine the atomic positions,
and determine their optimal heights about the Pt(111) surface,
we turn to density functional theory (DFT) calculations.
DFT calculations were performed using the code\cite{disclaim}
VASP.\cite{Kresse96} The PBEsol GGA exchange-correlation functional\cite{Perdew08}
was used, and the plane-wave cutoff was 500 eV.  Calculations were performed
using one k-point, at the origin.
A previous hybrid density functional theory study of 
Ti$_2$O$_3$ in the corundum structure shows a spin-paired 
electronic ground state.\cite{Guo12} 
To reproduce these results as closely as
possible at lower computational cost, we ran spin-unpolarized 
GGA+U calculations, after tuning the U parameters for Ti and O 
to reproduce the volume and bandgap of corundum Ti$_2$O$_3$ as 
accurately as possible (Ti U = 2.35 eV; O U = 0.30 eV).

 Quasicrystalline structures are not compatible with
periodic boundary conditions.  To investigate the stability
of our model, we turn instead to periodic approximants
to the quasicrystalline structure.  These periodic approximants
have the same local structures as the quasicrystals.
One such approximate has a unit cell 25.6~\AA~on an edge
with $\gamma=$ 120\degrees.  We created an approximate
to the G\"ahler $C_a$ tiling and decorated it as shown as
Figure 1.   We placed this structure above a model Pt-111 trilayer,
strained to have the same periodicity as the approximant.
The lowest Pt layer had all atomic coordinates fixed, the next layer
z was allowed to vary, and the topmost layer all atomic coordinates were allowed
to move.  A vacuum layer was created between the structure and periodic images by using a  
periodicity of 20~\AA~in the z-direction.  
The structure was relaxed until all forces were relaxed to less than 0.1 eV/~\AA.
An additional, square, approximant of side length 49.4~\AA~was created
and relaxed on a fixed Pt monolayer until all forces were less than 0.1 eV/~\AA.

The relaxed structures are shown in \fig{relax}(b-d).
The structures retain their ideal tile geometries quite well.
The most significant distortion is the clockwise or
counterclockwise rotation of 3 O around a Ti that frequently
occurs.  This rotation is often driven by the formation of
more favorable O-Ti-O bond angles for the Ti in the thin rhombs; 
{\it ab initio} molecular dynamics reveal frequent libration
of TiO$_3$ units at room temperature.
The average heights above the topmost Pt layer in the
relaxed structure are 3.1~\AA, 2.2~\AA, and 3.1~\AA~for
Ba, Ti, and O, respectively.  
Ti-O distances range from 1.80~\AA~to 1.95~\AA.
The rumpling of the Ti-O network (\fig{relax}(c)) 
is driven by transfer of electrons from Ba to Pt; the negative surface
charge on the Pt layer then attracts the positive Ti ions
and repels the negative O ions.\cite{Wu15}


\begin{figure}
\includegraphics[width=236pt]{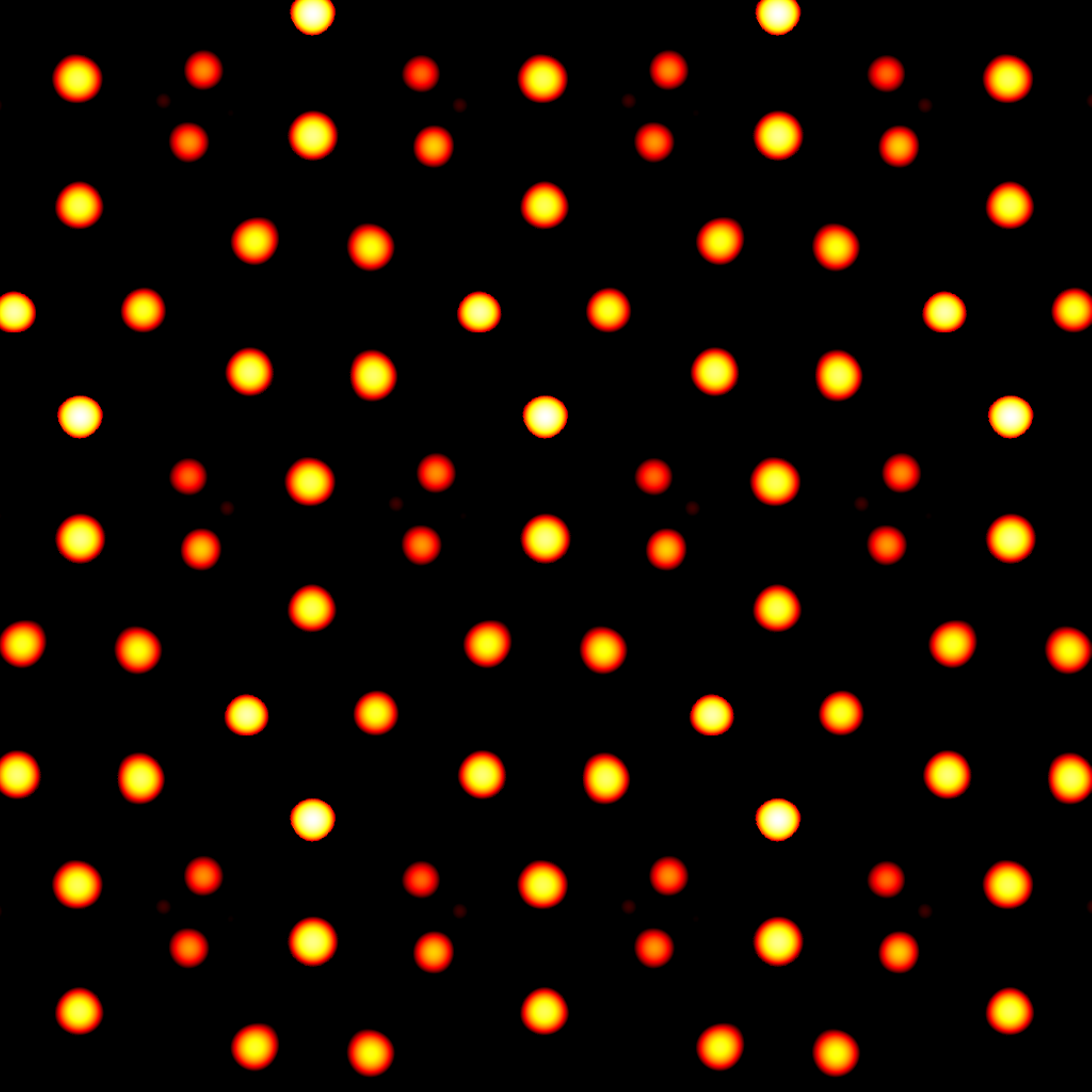}
\caption{Simulated scanning tunneling microscopy (STM) topograph 
of the quasicrystal approximant shown in \fig{relax}(b). Simulated
bias voltage $-0.15$ eV, charge density 0.1 e nm$^{-3}$,
and  topographic range 1~\AA.}
\label{fig:stm}
\end{figure}

STM topographs were simulated using the 
Tersoff-Hanann approximation.\cite{Tersoff85}
In this approximation, measuring a contour of constant current is equivalent to 
measuring a contour of constant electron density, projected 
in the energy range between $E_F$ and $E_F + e V$, with 
$E_F$ the Fermi level and $V$ the bias voltage.
The projected electron density was calculated via DFT,
starting with the DFT-relaxed structure in \fig{relax}(b).
The result is shown in \fig{stm}.
By comparison with the atomic structure, we see that
only the Ba atoms are visible and that the Ti-O network is invisible.
The simulated STM image strongly resembles the experimental
ones.\cite{Foerster12} The sizes of the simulated bright protrusions 
match even better with experiment if the computed projected charge density 
is convoluted by an in-plane factor  ${\rm exp}(-\rho^2/(2 \sigma^2))$, $\sigma$  =
1.0~\AA, to mimic experimental resolution.

\begin{figure}
\includegraphics[width=236pt]{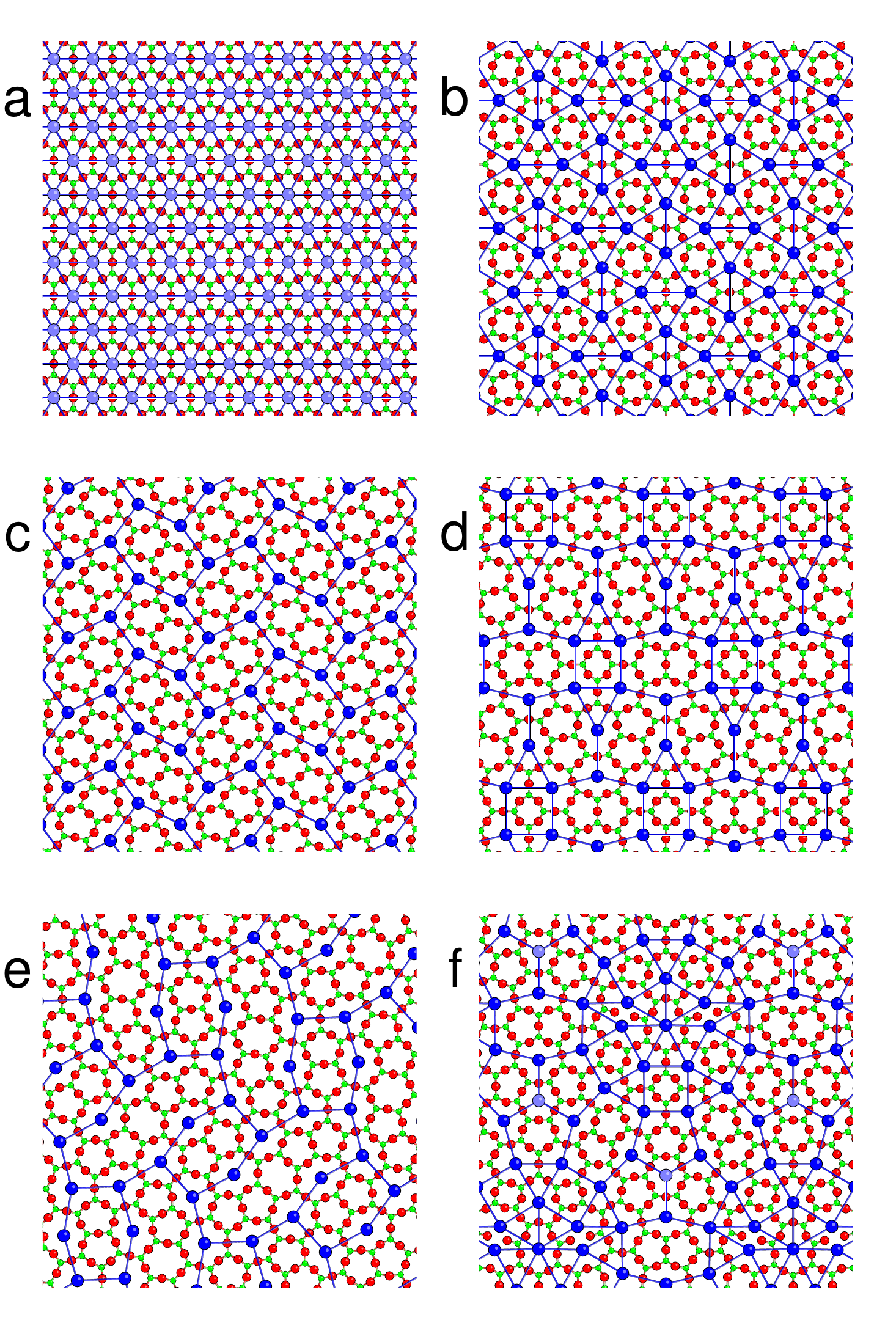}
\caption{Structure models for periodic crystalline thin film Ba-Ti-O.
(a) Structure model proposed by Wu {\it et al.}\cite{Wu15}
for thin film Ba-Ti-O on Au-111. Light blue color indicates
partial Ba occupancy.  (b-e): Structure models proposed in
present work for various periodic crystalline Ba-Ti-O structures 
observed by F\"orster {\it et al.}\cite{Foerster12}:
(b) Y-rows; (c) Six-network; (d) Double-Y; (e) Six-tiling;
(f) Wagon wheel.  There is a partially occupied Ba site shown
in light blue.}
\label{fig:cryst}
\end{figure}

 The presence of three-fold coordinated Ti atoms in the
quasicrystal Ba-Ti-O structure model along with Ti$_n$O$_n$
polygons with $n =$ 4, 5, and 7 is reminiscent of threefold
coordinated network structure observed in other two-dimensional
systems such as graphene\cite{Kotakoski11} and 
bilayer SiO$_2$.\cite{Huang12,Bjorkman13}  
The Ti-O-Ti links in the Ba-Ti-O system are analogous to 
the C-C bonds in graphene.  The Ba-Ti-O counterpart to the 
ideal graphene structure is shown in \fig{cryst}(a).  
This structure has actually been observed by Wu {\it et al.}\cite{Wu15}, 
from thin film Ba-Ti-O on a Au-111 surface,
although the Ba site is not fully occupied.

By assuming that the bright protrusions seen in the STM images of
the {\em periodic} Ba-Ti-O thin film structures observed by F\"orster 
{\it et al.}\cite{Foerster12} were Ba, and that the Ti and O formed a 
Ti-O network with threefold coordinated Ti, we were able to devise a 
structure model for each case (\fig{cryst}).
A bright spot that is sometimes present and sometimes absent in
the STM image of the ``wagon wheel" structure (\fig{cryst}(f))
is easily interpreted as partial Ba occupancy of a Ti$_6$O$_6$
pore.  The same kinds of tiles that occur in the quasicrystal model,
with the same decorations, occur in other structures, but 
several additional tiles are seen, each with a characteristic 
decoration: a thin hexagon, a pentagon, and
an elongated curved decagon.  The ``Y-rows" and, arguably, the
``Wagon wheel" structures are periodic approximants to the
quasicrystal, but the structures in \fig{cryst}(c-e) are unrelated to 
the quasicrystal.
The Ba:Ti ratios of the periodic crystalline Ba-Ti-O structures in
\fig{cryst} range from 3:13 $\approx$ 0.231 for the ``Six-tiling"
to 1:3 $\approx$ 0.333 for the ``Y-rows" structure\cite{footnote1}
to a hypothetical maximum of 0.5 for \fig{cryst}(a) with full Ba occupation.
The relatively high Ba:Ti ratio of 0.37 for the quasicrystal model
described earlier is due to the predominance of tiles with high
Ba:Ti ratios.

 With such a rich variety of observed structures, it becomes
possible to elucidate their common features.
Some observations: (1) The Ba-Ti-O systems show Ti$_n$O$_n$
rings with $n$ = 4, 5, 6, and 7, but not $n$ = 8 or greater; (2) The
$n$ = 7 rings are very common; and the $n$ = 6 rings relatively
rare; (3)  The 7-rings always share a Ti-O-Ti edge with two
or more other 7-rings to create a larger-scale network that
can be represented as a tiling; (4) The 7-rings are always
occupied by a Ba ion, the 6-rings partially occupied, and
smaller rings empty.

 The local structure associated with the rhomb in the
quasicrystal, its approximants, and the wagon wheel structure
can be viewed as the fusion of two Ti$_7$O$_7$-type
pores to make a dumbbell-shaped pore.  The two oxygen atoms that 
are coordinated with only one Ti serve as a buffer against Coulomb 
repulsion of the relatively close Ba-Ba pair.  
An analogous three-dimensional dumbbell-shaped pore was
recently found in a zeolite
structure,\cite{Smeets14} and the best structural model
in that case similarly had oxygen atoms inside the pore that
were bound to only one Si each, instead of the normal two.
The analogy with zeolites also suggests one possible application
of the structures proposed here: if freestanding
monolayers with the structures of the Ti-O networks shown here 
could be created, they could act as ultrathin filters with a uniform pore
size for gas separation, etc.

It remains an open question how the quasicrystal structure of 
Ba-Ti-O is stabilized relative to approximant and non-approximant
arrangements that can also occur.  The fact that these structures
form at high temperature suggests that entropy (vibrational and
perhaps tiling\cite{Elser85}) may play a key role.
In any case, the existence of detailed structure models for all
these structures is a first step toward solving this problem.

The authors thank Terrell A. Vanderah (NIST) for helpful discussions.
M. M. was supported by Slovak grants VEGA 2/0189/14 and APVV-0076-11.
C. L. H. received support from DOE grant DE-FG02-89ER45405.

\begin{thebibliography}{20}
\bibitem[{dis()}]{disclaim}
\bibinfo{note}{Certain commercial software is identified in this paper to
  adequately describe the methodology used. Such identification does not imply
  recommendation or endorsement by the National Institute of Standards and
  Technology, nor does it imply that the software identified is necessarily the
  best available for the purpose.}
\bibitem[{foo()}]{footnote1}
\bibinfo{note}{The arrangement of Ba and Ti in our model for the ``Six-network"
  structure is equivalent to the arrangement of Y and B in one layer of the
  YCrB$_4$ structure,\cite{Kuzma70} suggesting layered metal borides as a bulk
  system in which the structures presented in this paper might be achieved.}
\end{thebibliography}

\end{document}